
\tolerance=100000
\magnification=1200
\baselineskip=20pt

\font\title=cmbx10 scaled \magstep2

{\centerline{\bf \title Lattice Magnetic Walks}}

\vskip 0.25truein

{\centerline{Thomas Blum$^{\dagger}$}}

\vskip 0.05truein

{\baselineskip=12pt
{\centerline{Department of Physics and Astronomy}}
{\centerline{University of Pittsburgh}}
{\centerline{Pittsburgh, PA 15260}}
{\centerline{USA}}
}

\vskip 0.1 truein

{\centerline{Yonathan Shapir}}

\vskip 0.05truein

{\baselineskip=12pt
{\centerline{Department of Physics and Astronomy}}
{\centerline{University of Rochester}}
{\centerline{Rochester, NY 14627-0011}}
{\centerline{USA}}
}

\vskip 0.5truein

{\centerline{Abstract}}

Sums of walks for charged particles ({\it e.g.} Hofstadter
electrons) on a square lattice in the presence of a magnetic
field are evaluated.
\ Returning loops are systematically added to directed paths to
obtain the unrestricted propagators.
\ Expressions are obtained for special values of the magnetic
flux-per-plaquette commensurate with the flux quantum.
\ For commensurate and incommensurate values of the flux, the
addition of small returning loops does not affect the general
features found earlier for directed paths.
\ Lattice Green's functions are also obtained for staggered flux
configurations encountered in models of high-$T_c$ superconductors.

\vfill
\eject

\noindent {\bf I. Introduction}

\vskip 5pt

The problem of electrons confined to a lattice in the presence
of a transverse magnetic field is a classic problem of
mathematical physics with many applications in condensed-matter
physics.
\ The early works of Hofstadter [1], Wannier [2], and Azbel [3]
focused on the spectral properties of non-interacting electrons
as a function of the energy and the parameter $\alpha=\phi/\phi_o$
where $\phi$ is the magnetic flux-per-plaquette and $\phi_o=hc/e$
is the flux quantum.
\ For commensurate $\alpha=p/q$ the spectrum has $q$ subbands with
special scaling properties as $q$ grows larger, eventually turning
into a Cantor set for irrational $\alpha$.
\ Hence, the behavior of the spectrum and the wave function is very
sensitive to the exact value of $\alpha$ and the energy $E$; (the
intricate structure of the bands as a function of $\alpha$ and $E$
has been dubbed the ``butterfly").

Lately the renewed interest in interacting electrons in two
dimensions, in the context of the Quantum Hall Effect and various
theories for high-$T_c$ superconducting materials for instance,
has led to more recent investigations of this general problem.
\ Over time, these works have yielded an extraordinary
richness using a variety of techniques including path integrals [4],
renormalization group [5], sophisticated algebraic methods [6]
closely related to ``quantum groups," and so on.

More recently the study of non-interacting electrons in a
magnetic field has been approached from a new direction --- that
of ``localized" wave functions. [7-10]
\ If the energy of the electron does not lie within a quasiband, its
eigenfunction is not a bulk (extended) eigenstate.
\ However, such a wave function may be localized at inhomogeneities,
such as the edge of the lattice ({\it i.e.}, surface gap states) or at
isolated impurities in an otherwise ordered lattice.
\ These states decay exponentially as one moves away from the
inhomogeneity into the bulk.
\ The effect of the magnetic field on this exponential decay (and
particularly the associated ``localization length") has been studied
in the so-called ``directed-paths approximation." [7,10]

These calculations are based on the transfer matrix approach
(reviewed below) which is applicable only in this approximation.
\ While it can be shown that this approximation yields a good
description in the extremely localized limit in the absence of any
field, the question is more delicate when the magnetic field is
present.
\ Indeed, to the next order, paths with small returning loops should be
added.
\ In the absence of a magnetic field, this inclusion leads to a simple
renormalization of the ``localization length."
\ On the other hand, one finds already in the directed-paths
approximation that in the presence of the field, the combined
effects of lattice periodicity and the magnetic flux yield a
complex interference pattern with a very sensitive dependence of
the localization length upon the flux. [10]
\ Therefore, the effects of including even small returning loops
with their enclosed flux are unpredictable and potentially may be
very drastic.

Including the returning loops, as explained below, becomes even more
important if the energy to begin with is not extremely far from
the band edge and the wave functions are consequently only
moderately localized.
\ We thus set out in this work to go beyond the directed-paths
approximation and to calculate systematically the full Green's
functions which include the sum over all paths.
\ These will be given as formal expressions which may be expanded
systematically in various ways.
\ Notable among these is the expansion in terms of the path length
starting with the directed (the shortest) ones; it reveals that the
generic features found in the directed-paths limit are not drastically
changed by the small returning loops.
\ In this paper we restrict ourselves to the square lattice,
but using this approach, the sums for other lattices may be
calculated as well.
\ The expression we derive may also be useful to analyze properties
of Hofstadter electrons within the band of bulk eigenstates.

Nonuniform flux configurations have been studied recently in the
context of theories of strongly interacting electrons relevant
to high-$T_c$ cuprates.
\ The most important among them are the staggered flux
configurations [11].
\ We consequently obtain here the lattice Green's function
for the square lattice with staggered flux as well.

The layout of this paper is as follows.
\ In section 2, we review the model and the ``directed-paths-only''
transfer-matrix approach.
\ The heart of the paper is section 3, where we consider the
addition of ``backward excursions'' to the transfer-matrix
method.
\ Here, we find recursion relations and other expressions
for the various quantities involved.
\ In section 4, we examine some special cases thereof.
\ Section 5 demonstrates how the full Green's function can be
obtained from the quantities derived in sections 3 and 4.
\ Section 6 considers the staggered flux configuration, and
section 7 contains a discussion of the results.
\ We also include an appendix, listing some additional
results.

\vskip 10pt

\noindent {\bf II. The Model and the Approach}

\vskip 5pt

One can model non-interacting electrons confined to a square
lattice with a perpendicularly-applied magnetic field by the
following tight-binding hamiltonian:
$$H \ =\ \epsilon \sum_i a_i^{\dagger}a_i + V\sum_{\langle ij\rangle}
a_i^{\dagger}a_j {\rm e}^{i\gamma_{ij}}, \eqno(2.1)$$
where $\epsilon$ is the on-site energy and $V$ measures the overlap
of wave functions centered at neighboring sites.
\ The energy spectrum is the same as the one obtained using the
``Peierls substitution,'' which replaces $\hbar \vec k$ with
$\vec p -e\vec A/c$ in the dispersion relation obtained from the
tight-binding approximation in the non-magnetic case.
\ The phase $\gamma_{ij}=-\gamma_{ji}$ denotes the Aharonov-Bohm
phase an electron acquires as it moves from site $i$ to site $j$;
physically, it corresponds the line integral of the vector potential
from $i$ to $j$.
\ Some freedom exists in the choice of gauge:
\ To characterize a constant magnetic field, one selects the phases
such that the directional sum of the phases $\gamma_{ij}$ around each
elementary plaquette is $2 \pi \phi / \phi_o$ where $\phi$ is the
magnetic flux-per-plaquette and $\phi_o$ the flux quantum ($hc/e$).

Many of the physically interesting quantities are contained in
or are derivable from the Green's function associated with eq.
(2.1).
\ One can express the (real-space) Green's function ${\cal G}
(i,f,E,\phi)$ between sites $i$ and $f$ for an electron of energy
$E$ as a sum over paths connecting the two sites:
$${\cal G}(i,f;E,\phi) \ =\ \sum_{paths} \ \prod_{\langle jk \rangle }
\ {V \ {\rm exp}\bigl\{i\gamma_{jk}\bigr\}
\over E-\epsilon}~, \eqno(2.2)$$
where the product is over steps comprising a given path.
\ Each path carries a weight $v^L$, where $v={V \over E-\epsilon}$
and $L$ is the length of the path.
\ Since the total number of paths on a square lattice grows as $4^L$,
the sum should converge provided $v<{1 \over 4}$.
\ The model has a phase transition from localized to extended states
occurring at $v={1 \over 4}$. [5]

In the strong localization regime ($v \ll {1 \over 4}$),
the electronic wave function decays quite rapidly as one moves away
from some initial position.
\ In this case, ${\cal G}(i,f;E,\phi)$ is dominated by the shortest
paths --- those ``directly'' connecting the initial and final sites.
[12]
\ Note that on a lattice there exists in general more than one
shortest path connecting two sites.
\ The presence of a magnetic field gives rise to interference among
various constituent paths.
\ Since it is the interference phenomena that interests us and
since the difference in the phases accumulated between two paths
is proportional to the area they enclose, we find it convenient to
study the geometry in which directed paths can enclose the largest
possible areas.
\ For this reason, we begin our study with paths directed along the
diagonal of a square lattice; we will call the directed axis $t$ and
the transverse axis $x$.
\ (See Figure 1.)
\ Summation of paths directed along a lattice axis (rather than
the diagonal) is also possible but slightly less convenient.

One can employ the transfer-matrix approach to sum over directed
paths. [7]
\ If the vector $|v_{t-1}\rangle$ represents a walk or walkers
at column $t-1$, then the matrix $T(t, \phi)$ operating on
$|v_{t-1}\rangle$
$$|v_t \rangle \ =\ T(t,\phi) ~|v_{t-1}\rangle \eqno(2.3)$$
provides all possible extensions of these walks by one forward
step and weights these possibilities appropriately.
\ Consecutive operation by transfer matrices hence sums over
all possible walks
$$|v_t\rangle \ =\ T(t,\phi)\ T(t-1,\phi) \
\ldots T(1, \phi) ~|v_0\rangle. \eqno(2.4)$$
Summing over paths weighted by different phase factors yields
the desired interference effects.

The transfer matrix connecting column $t-1$ to column $t$
is:
$$T(t,\phi)\ =\ v \left[\matrix{0 & e^{-it \phi \pi} & 0 & ...
& 0 & 0 & e^{it \phi \pi} \cr
e^{it \phi \pi} & 0 & e^{-it \phi \pi} & ... & 0 & 0 & 0 \cr
0 & e^{it \phi \pi} & 0 & ... & 0 & 0 & 0 \cr
 \vdots & \vdots & \vdots &  & \vdots & \vdots & \vdots \cr
0 & 0 & 0 & ... & 0 & e^{-it \phi \pi} & 0 \cr
0 & 0 & 0 & ... & e^{it \phi \pi} & 0 & e^{-it \phi \pi} \cr
e^{-it \phi \pi} & 0 & 0 & ... & 0 & e^{it \phi \pi} & 0 \cr
}\right],\eqno(2.5)$$
where the bonds are viewed as being directed from lower to higher
$t$.
\ We exploit the gauge degree-of-freedom to furnish a transfer
matrix that depends only on $t$ (and not on $x$).
\ This choice is called the diagonal staggered gauge [7].
\ As required, summing the phases as one proceeds clockwise
around any elementary plaquette results in $2 \pi \phi$;
we have set $\phi_o=hc/e =1$ for convenience.

Note that in addition we have applied periodic boundary
conditions along the $x$ axis.
\ (However, since we approach the problem from the localization
side, we may always take the lateral extent $x_{max}$ of the
contributing paths to be smaller than the vertical size of the
system $L_{\perp}$, and so
the sum over paths will not be sensitive to the boundary conditions.
[13])
\ With periodic boundary conditions, the eigenvectors of all $T$'s
are plane waves:
$$| v_k \rangle \ =\ {1 \over \sqrt{L_{\perp}}}
                     \left[ \matrix{ e^{ik_{\perp}} \cr
                                     e^{i2k_{\perp}} \cr
                                     e^{i3k_{\perp}} \cr
                                     \vdots \cr
                                     e^{iL_{\perp} k_{\perp}} \cr}
\right]. \eqno(2.6)$$
The corresponding eigenvalues are $\lambda_{k_{\perp}}(t,\phi)
=2v \cos (k_{\perp}-t \phi \pi)$.
\ The transverse momenta take on values $k_{\perp}=2 \pi m/L_{\perp}$,
where $L_{\perp}$ is the vertical size of the lattice and $m=0,1,...,
L_{\perp}-1$.
\ The eigenvalues might also be called $G_{k_{\perp}}^{dir}
(t-1,t,\phi)$ for they correspond to the Green's function for
directed walks (Fourier transformed in the transverse direction)
joining columns $t-1$ and $t$.

Note that just as in the continuum model, one finds plane waves in
the ``transverse'' direction, leaving an effective one-dimensional
problem (an harmonic oscillator in the continuum case).
\ The reduction to a one-dimensional problem does not depend
on the restriction to directed walks since the same transfer
matrix also yields a ``backward'' step from column $t$ to $t-1$.
\ Henceforth, we need only consider eigenvalues, as all matrices
of interest are simultaneously diagonalizable by the plane-wave
eigenvectors.

In the presence of a magnetic field, the exponential decay of the
electron wave function away from its central position is modulated
by a very intricate interference pattern.
\ In the strongly localized regime, many of its properties can be
uncovered by investigating the product obtained from the
directed paths only
$$G^{dir}_{k_{\perp}}(0,T_o,\phi) \ =\ \prod_{t=0}^{T_o-1}
G^{dir}_{k_{\perp}}(t,t+1,\phi) \ =\
\prod_{t=1}^{T_o} \lambda_{k_{\perp}}(t,\phi)
\ =\ \prod_{t=1}^{T_o} \Bigl[ 2v \ C_t \Bigr], \eqno(2.7a) $$
where for convenience, we have introduced the notation
$$C_t=\cos
(k_{\perp}-{t \phi \pi}); \eqno(2.7b)$$
suppressing the $\phi$ and $k_{\perp}$
dependences as most of the calculations considered herein
involve a single $\phi$ and a single $k_{\perp}$.
\ $G^{dir}_{k_{\perp}}(0,T_o,\phi)$ supplies the Green's function
for directed walks that begin at the origin and end in column $T_o$.
\ The properties of this product have been the focus of a
recent investigation. [10]
\ The transition amplitude between two sites, say $(0,0)$ and
$(x_o,T_o)$, may be obtained from them through:
$$G(0,0;x_o,T_o) \ =\ {1 \over L_{\perp}}
\sum_{k_{\perp}} G^{dir}_{k_{\perp}}
(0,T_o;\phi) {\rm e}^{-ik_{\perp}x_o} . \eqno(2.8)$$

In the absence of any field, the directed Green's function is simply
a product of cosines:
$$G_{k_{\perp}}^{dir}(0,T_o,0)\ =\ \Bigl[2v ~\cos (k_{\perp})
\Bigr]^{T_o}. \eqno(2.9)$$
When the applied flux-per-plaquette is a rational multiple
of the flux quantum ({\it i.e.} ${\phi \over \phi_o} ={p \over q}$
with $p$ and $q$ relatively prime integers), the
interference-induced modulation is periodic; in fact, the pattern
on sites of a {\it superlattice} with lattice constant $q$ times
larger than the original lattice mimics precisely the simpler decay
pattern found in the absence of an applied field. [7-10]
\ If we restrict our attention to sites on the superlattice in the
commensurate case, then
$$G_{k_{\perp}}^{dir}(0,T_o,{\textstyle{p \over q}})
\ =\ \Biggl[~\prod_{t=1}^{q}
2v \ C_t ~\Biggr]^{T_o/q}. \eqno(2.10)$$
One can then use the following property:
$$\prod_{t=1}^{q} C_t
\ =\ \cases{2^{-q+1} \cos (qk_{\perp} ),
&\ \ if $q$ is odd; \cr
2^{-q+1}\sin (qk_{\perp} ),
&\ \ if $q$ is even;\cr}\eqno(2.11)$$
to see the relation between the commensurate case on the superlattice
and the nonmagnetic case.
\ Note that in the commensurate case,
$v^{-T_o}G_{k_{\perp}=0}^{dir}(0,T_o,{\textstyle{p \over q}})
\propto 2^{T_o/q}$.

The work of Fishman, Shapir and Wang [10] addresses the behavior of
this quantity in the incommensurate case ({\it i.e.} when $q
\rightarrow \infty$).
\ In this case, the structure becomes aperiodic.
\ For generic irrational, it has been found that ${\rm ln}
|v^{-T_o}G^{dir}|$ increases as $[{\rm ln} (T_o)]^2$; while for an
algebraic irrational ${\rm ln} |v^{-T_o}G^{dir}|$ increases as
${\rm ln}(T_o)$.

\vskip 10pt

\noindent {\bf III. Allowing Backward Excursions}

\vskip 5pt

The main motivation for the present work is to consider the
consequences of lifting the ``directed paths only'' restriction
from the previous analysis, thereby broadening its scope
beyond the strongly localized regime.
\ We have opted to pursue the transfer-matrix method; however,
this choice requires reconciling that approach, so ideally suited
to directed paths, with our current concern of including returning
loops.
\ We explain first how the formal expressions can be obtained in
an elegant and minimal way; we then proceed to a more detailed
derivation which starts from the directed paths and adds systematically
longer and longer returning loops.
\ The expressions gained by this latter method will form the
basis for our conclusions on the effects of adding small loops to
fully directed paths.
\ With these ends in mind, let us begin our investigation of
``directed paths with backward excursions,'' which we define next.

Let $\tilde \lambda_{k_{\perp}}(t,\phi)$ be the eigenvalue of
a transfer matrix $\tilde T(t,\phi)$ which allows any amount of
backward excursion, that is, any number of loops of any length
starting from $t-1$ and ending at $t$ --- provided only the last step
bridges columns $t-1$ and $t$.
\ (See, for example, Figure 1.)
\ The eigenvalue $\tilde \lambda_{k_{\perp}}(t, \phi)$ might also
be called $G^{b.e.}_{k_{\perp}}(t-1,t,\phi)$ for it denotes
the Green's function linking columns $t-1$ and $t$ that admits
``backward excursions.''
\ The full, unrestricted Green's function can then be
calculated from $G^{b.e.}_{k_{\perp}}(s,t,\phi)$ as will be
shown in section V.

In view of the fact that all of the terms comprising
$\tilde \lambda_{k_{\perp}}(t,\phi)$ contain a
factor $2v C_t$ corresponding to the last step, it
is convenient to define $\hat \lambda_{k_{\perp}}
(t,\phi)= \tilde \lambda_{k_{\perp}}(t,\phi)/2vC_t$
with this last piece stripped off.
\ Note that the walks contributing to $\hat \lambda_{k_{\perp}}
(t,\phi)$ begin and end in column $t-1$, never reaching column $t$.
\ Again we could use the Green's function notation $\hat
\lambda_{k_{\perp}}(t,\phi)=G_{k_{\perp}}^{b.e.}(t-1,t-1;\phi)$
which is more descriptive but a bit more cumbersome, and so
we will put off its use until section V.

We have found that $\hat \lambda_{k_{\perp}}(t,\phi)$ obeys the
recursion relation:
$$\eqalignno{
\hat \lambda_{k_{\perp}}(t,\phi) \ &=\
1 \ +\  \Bigl[2v C_{t-1} \Bigr]^2\  \hat
\lambda_{k_{\perp}}(t-1,\phi) \cr
&\ \ \ \ \ \
+\ \biggl\{ \Bigl[ 2v C_{t-1} \Bigr]^2 \
\hat \lambda_{k_{\perp}}(t-1,\phi) \biggr\}^2
\ +\  .... &(3.1a)\cr }$$
This relation can be understood in the following way:
\ In the expansion, the first term on the right-hand side
(the $1$) indicates the option of making no backward steps.
\ In the second term, one of the $2v C_{t-1}$'s denotes a step
back to column $t-2$, then the $\hat \lambda_{k_{\perp}}
(t-1, \phi)$ designates any amount of backward excursion beyond
$t-2$ which eventually returns to column $t-2$ (including again
the option of no excursion at all) and the second $2v C_{t-1}$
returns the walker to column $t-1$.
\ Finally this whole process might be repeated any number of
times as indicated by the expansion.
\ The expansion is readily summed (at least formally) as a
geometric series.
$$
\hat \lambda_{k_{\perp}}(t,\phi) \ = \
\biggl\{ 1 \ -\  4v^2 C_{t-1}^2
\ \hat \lambda_{k_{\perp}}(t-1, \phi)\biggr\}^{-1},
\eqno(3.1b) $$

Next, consecutive $\hat \lambda$'s
$\bigl[ \hat \lambda_{k_{\perp}}(t-1,\phi), ~\hat
\lambda_{k_{\perp}}(t-2,\phi),\ldots\bigr]$
can be substituted into eq. (3.1b) to arrive at the
following continued-fraction expression:
$$\hat \lambda_{k_{\perp}}(t,\phi) \ =\
{1 \over \displaystyle \ 1\ -\
{\strut 4v^2 C_{t-1}^2 \over \displaystyle \ 1\ -\
{\strut 4v^2 C_{t-2}^2 \over \displaystyle \ 1\ -\
{\strut 4v^2 C_{t-3}^2 \over etc.}}}}. \eqno(3.2)$$
Terminating this continued fraction at the $C_{t-s}^2$ term
and re-expressing $\hat \lambda$ as a ratio of polynomials,
one finds that the numerator and denominator have essentially the
same form except that the terms containing $C_{t-1}$ are absent in
the numerator.
\ More specifically, the procedure leads to:
$$\hat \lambda_{k_{\perp}} (t,\phi) \ = \
{}~{D_{k_{\perp}} (t,\phi;2,s)
\over D_{k_{\perp}} (t,\phi;1,s)} \ +\  O(v^{2s+2}) ,
\eqno(3.3a)$$
where
$$\eqalignno{
D_{k_{\perp}}(t, \phi;r,s) \ =\ &1 \ -\
4v^2 \sum_{j=r}^{s} C_{t-j}^2
\ +\ 16 v^4 \sum_{j=r}\ \sum_{k=j+2}^{s}C_{t-j}^2 C_{t-k}^2 \cr
&-\ 64v^6 \sum_{j=r}\ \sum_{k=j+2}\
\sum_{l=k+2}^{s} C_{t-j}^2 C_{t-k}^2 C_{t-l}^2 \ + \ ....
&(3.3b) \cr}$$

The problem has thus been reduced to that of understanding the
properties of this particular function $D_{k_{\perp}}(
t,\phi;r,s)$.
\ Some of its more obvious and useful properties are:
$$\eqalignno{
D_{k_{\perp}}(t,\phi;r,s) \ =\ &
{}~D_{k_{\perp}}(t+n,\phi;r+n,s+n), &(3.4a)\cr
D_{k_{\perp}}(t,\phi;r,s) \ =\ &
{}~D_{k_{\perp}}(t,\phi;r+1,s)~-~4v^2C_{t-r}^2
D_{k_{\perp}}(t,\phi;r+2,s), &(3.4b)\cr
D_{k_{\perp}}(t,\phi;r,s) \ =\ &
{}~D_{k_{\perp}}(t,\phi;r,s-1)~-~4v^2C_{t-s}^2
D_{k_{\perp}}(t,\phi;r,s-2). &(3.4c)\cr }$$

Much is encoded in eq. (3.3).
\ However, since our motivation is to extend the analysis of fully
directed walks to include returning loops, we present a methodical
derivation which does just that.
\ Along the way useful expansions for $\hat \lambda_{k_{\perp}}
(t,\phi)$ are obtained.

Returning then to the directed paths, a first step toward including
returning loops would permit in addition to one forward step
connecting columns $t-1$ and $t$, one backward step followed by
two forward steps.
\ The modified eigenvalues $\lambda^m_{k_{\perp}}(t,\phi)$
become:
$$\lambda^m_{k_{\perp}}(t,\phi) \ =\ \lambda_{k_{\perp}}(t,\phi)
\ +\ \bigl[\lambda_{k_{\perp}}(t-1, \phi)\bigr]^2
{}~\lambda_{k_{\perp}}(t, \phi). \eqno(3.5)$$
Let us analyze the effect of this addition on the Green's function
for one plaquette on the superlattice in the case of commensurate
flux.
\ As in eq. (2.7a) for directed paths, the Green's function is
a product of these eigenvalues, and this product can be separated
into two products as follows:
$$
G^m_{k_{\perp}}\Bigl(0,q, {\textstyle {p \over q}}
\Bigr) \ =\
\Biggl[\prod_{t=1}^{q} 2vC_t \Biggr]
\times \Biggl[\prod_{t=1}^{q} \Bigl( 1 +4v^2
C^2_{t-1} \Bigr) \Biggr]. \eqno(3.6a) $$
The first product is $G^{dir}_{k_{\perp}}(0,q,
{\textstyle{p \over q}})$ which we considered in eqs. (2.9)
and (2.10); the second product is also readily handled;
together they yield:
$$ G^m_{k_{\perp}}\Bigl(0,q, {\textstyle {p \over q}}
\Bigr) \ =\
G^{dir}_{k_{\perp}}\Bigl(0,q, {\textstyle {p \over q}} \Bigr)
\times (1 -\alpha)^{-2q}
\Bigl[ 1 - (-1)^q 2 \alpha^q \cos (k_{\perp}q) +
\alpha^{2q} \Bigr], \eqno(3.6b) $$
where
$$\alpha \ =\ {1+2v^2 - \sqrt{1+4v^2} \over 2v^2} \ \approx
\ v^2 -2v^4. \eqno(3.6c)$$
{}From this expression, one can see that the main effect is a
renormalization of $v$ without any dependence on $\phi$.
\ The only structural change arises from the $\cos (k_{\perp}q)$
term; however, it is of the order $v^{2q}$ times smaller than
the leading behavior.
\ Next, let us consider the inclusion of longer backward excursions.

As one moves further out of the strongly localized regime, one should
account for paths with increasingly longer backward excursions,
culminating in $\tilde \lambda_{k_{\perp}}(t, \phi)$ which
consists of contributions of all lengths.
$\tilde \lambda_{k_{\perp}}(t,\phi)$ can be written as:
$$\tilde \lambda_{k_{\perp}}(t,\phi) \ =\
\sum_{j=0}^{\infty} ~\tilde \lambda^{(j)}_{k_{\perp}}(t, \phi),
\eqno(3.7)$$
where $\tilde \lambda^{(j)}_{k_{\perp}}(t, \phi)$ designates
the contribution of length $(2j+1)$ which therefore carries a
weight $(2v)^{2j+1}$.
\ The first three are:
$$\eqalignno{
\tilde \lambda_{k_{\perp}}^{(0)}(t,\phi) \ =\ &
(2v)C_t &(3.8a) \cr
\tilde \lambda_{k_{\perp}}^{(1)}(t,\phi) \ =\ &
(2v)^3C_tC_{t-1}^2 &(3.8b) \cr
\tilde \lambda_{k_{\perp}}^{(2)}(t,\phi) \ =\ &
(2v)^5 C_t \Bigl[C_{t-1}^4 + C_{t-1}^2C_{t-2}^2 \Bigr].
&(3.8c) \cr }$$
We provide expressions for $\tilde \lambda_{k_{\perp}}^{(j)}
(t,\phi)$ for $j=0,1,\ldots,5$ in the appendix.
\ They become cumbersome rather quickly as $j$ increases.
\ For $j \geq 1$, $\tilde \lambda_{k_{\perp}}^{(j)}(t, \phi)$
can be shown to obey the following recursion relation:
$$ \tilde \lambda_{k_{\perp}}^{(j)}(t, \phi) \ =\
\left[ ~\sum_{\ell=1}^j \tilde \lambda_{k_{\perp}}^{(j-\ell)}
(t, \phi) \ \tilde \lambda_{k_{\perp}}^{(\ell-1)}(t-1, \phi)
{}~\right] ~\tilde \lambda_{k_{\perp}}^{(0)}(t-1, \phi). \eqno(3.9)$$

Again we find it convenient to strip off the last step of
the walk and study $\hat \lambda_{k_{\perp}}(t, \phi)$.
After generating $\hat \lambda_{k_{\perp}}^{(j)}(t,\phi)$'s
from the recursion relation, we have observed that they could
be expressed in the following way:
$$\hat \lambda_{k_{\perp}}^{(j)}(t, \phi) \ =\
\sum_{\{n_i | {{
\sum_{i=1}^j}} n_i=j \}} \ \Biggl[ \ \prod_{i=1}^{\infty}
{n_i+n_{i+1}-1 \choose n_{i+1}}
\Bigl(4v^2 \ C_{t-i}^2 \Bigr)^{n_i} \ \Biggr] , \eqno(3.10a)$$
where $n_i=0,1,\ldots,j$ for $i=1,2,\ldots,j$ and where
$$ {n_i+n_{i+1}-1 \choose n_{i+1}}
\ =\ \cases{~1, &\ \ if $n_i=0$ and $n_{i+1}=0$; \cr
            ~0, &\ \ if $n_i=0$ and $n_{i+1}>0$; \cr
{}~{\displaystyle{(n_i+n_{i+1}-1)! \over (n_{i+1})!(n_i-1)!}},
&\ \ otherwise. \cr} \eqno(3.10b)$$
The restriction on the $n_i$'s in eq. (3.10a) ensures that the
contributing walks have the appropriate length.
\ Note first of all that each cosine term appears an even
number of times, making the round trip possible.
\ Moreover, the terms are ``connected;'' for instance, if
a term contains a factor $C_{t-3}$, it must also contain
factors $C_{t-2}$ and $C_{t-1}$.
\ Finally, the numerical component (the degeneracy) factorizes
into terms which depend only on the number of steps in two
neighboring columns ($n_i$ and $n_{i+1}$).

Expresson (3.10) may have certain practical advantages over eq. (3.3)
in that it selects out walks of a particular length and that it is
not written in the form of a quotient.
\ Nevertheless, we proceed now to use it in a second derivation of
eq. (3.3).
\ Performing the sum of $\hat \lambda_{k_{\perp}}^{(j)}(t, \phi)$
over $j$
$$
\hat \lambda_{k_{\perp}}(t,\phi) \ =\
\sum_{j=0}^{\infty} ~\hat \lambda_{k_{\perp}}^{(j)}(t,\phi)
\eqno(3.11a) $$
actually simplifies matters as it eliminates the restriction
on $n_i$'s, leading to
$$
\hat \lambda_{k_{\perp}}(t,\phi)
\ =\ \sum_{\{n_1,n_2,...\}} \ \Biggl[ \ \prod_{i=1}^{\infty}
{n_i+n_{i+1}-1 \choose n_{i+1}}
\Bigl(4v^2 C_{t-i}^2 \Bigr)^{n_i} \ \Biggr].
\eqno(3.11b)$$
Next carry out the sums over $n_i$ consecutively.
\ Using repeatedly the following sum:
$$\sum_{j=0} {j+k-1 \choose k}~x^j
\ =\ 1 \ +\ {x \over (1-x)^{k+1}}, \eqno(3.12)$$
leads to:
$$\eqalignno{
&\hat \lambda_{k_{\perp}}(t, \phi) \ =\
1 \ +\  {(2v)^2C_{t-1}^2 \over \bigl(1-4v^2C_{t-1}^2\bigr)}
\ +\ {(2v)^4 C_{t-1}^2 C_{t-2}^2 \over \bigl(1-4v^2C_{t-1}^2
\bigr) \bigl(1-4v^2C_{t-1}^2-4v^2C_{t-2}^2\bigr)}\cr
&\ \ \ \
+\ {(2v)^6 C_{t-1}^2 C_{t-2}^2 C_{t-3}^2 \over
\bigl(1-4v^2C_{t-1}^2-4v^2C_{t-2}^2\bigr)
\bigl(1-4v^2C_{t-1}^2-4v^2C_{t-2}^2 -4v^2C_{t-3}^2
+16v^4C_{t-1}^2C_{t-3}^2 \bigr)}\cr
&\ \ \ \ \ \ \ \ \ \ \ \ \ \ \
+ \ O(v^8). &(3.13)\cr }$$

Finally finding common denominators for the first two terms,
then the first three terms, etc., yields expressions like:
$$\eqalignno{
\hat \lambda_{k_{\perp}}(t, \phi) \ &=\ { 1-4v^2C_{t-2}^2
\over 1-4v^2C_{t-1}^2 -4v^2C_{t-2}^2} \ +\  O(v^6), \cr
\hat \lambda_{k_{\perp}}(t, \phi) \ &=\ { 1-4v^2C_{t-2}^2
-4v^2C_{t-3}^2 \over 1-4v^2C_{t-1}^2 -4v^2C_{t-2}^2 -4v^2C_{t-3}^2 +
16v^4C_{t-1}^2C_{t-3}^2} \ +\  O(v^8), &(3.14)\cr}$$
and so on.
\ Hence, one arrives at:
$$\hat \lambda_{k_{\perp}} (t, \phi) \ =\
{D_{k_{\perp}} (t,\phi;2,s)
\over D_{k_{\perp}} (t,\phi;1,s)}
\ + \ O(v^{2s+2}), \eqno(3.15)$$
which is identical to eq. (3.3).

\vskip 10pt

\noindent{\bf IV. Special Cases}

\vskip 5pt

In this section, we consider the evaluation of $\hat
\lambda_{k_{\perp}}(t,\phi)$ in three special cases: \
1)~when $v$ is small (which provides a straightforward
extension of the directed-paths-only analysis in the strongly
localized regime);
2)~when the magnetic flux-per-plaquette is a rational multiple
of the flux quantum;
and 3)~when $\phi$ is small (weak magnetic field, for which
one should be able to make some correspondence with the
continuum model).

\vskip 0.1truein

\noindent{\bf iv.a. When $v$ is small}

The first special case we shall treat is when $v$ is small.
\ Under such conditions, it is reasonable to view the various
quantities as power series in $v$.
\ We have already seen in the previous section that
$$\eqalignno{
\hat \lambda_{k_{\perp}}^{(0)}(t,\phi) \ =\ &
1 &(4.1a) \cr
\hat \lambda_{k_{\perp}}^{(1)}(t,\phi) \ =\ &
(2v)^2 C_{t-1}^2 &(4.1b) \cr
\hat \lambda_{k_{\perp}}^{(2)}(t,\phi) \ =\ &
(2v)^4 \Bigl[C_{t-1}^4 + C_{t-1}^2C_{t-2}^2 \Bigr],
&(4.1c) \cr
\hat \lambda_{k_{\perp}}^{(3)}(t,\phi) \ =\ &
(2v)^6 \Bigl[C_{t-1}^6 + 2C_{t-1}^4C_{t-2}^2
+ C_{t-1}^2C_{t-2}^4 + C_{t-1}^2C_{t-2}^2C_{t-3}^2
\Bigr], &(4.1d) \cr }$$
and so on.
\ From these, we see that $\hat \lambda_{k_{\perp}}(t,\phi)$
can be re-expressed as
$$\eqalignno{
\hat \lambda_{k_{\perp}}(t, \phi) \ =\
{\rm exp} \biggl\{&(2v)^2C_{t-1}^2 \biggr\}
\ +\ O(v^4).
&(4.2a) \cr
\hat \lambda_{k_{\perp}}(t, \phi) \ =\
{\rm exp} \biggl\{&(2v)^2C_{t-1}^2 + (2v)^4
\Bigl[ {\textstyle{1 \over 2}}C_{t-1}^4 +
C_{t-1}^2C_{t-2}^2 \Bigr] \biggr\} \ +\ O(v^6).
&(4.2b) \cr
\hat \lambda_{k_{\perp}}(t, \phi) \ =\
{\rm exp} \biggl\{&(2v)^2 C_{t-1}^2 + (2v)^4
\Bigl[ {\textstyle{1 \over 2}}C_{t-1}^4 +
C_{t-1}^2C_{t-2}^2 \Bigr] \cr
+&(2v)^6 \Bigl[{\textstyle{1 \over 3}}C_{t-1}^6 +
C_{t-1}^4C_{t-2}^2 + C_{t-1}^2C_{t-2}^4 +
C_{t-1}^2C_{t-2}^2C_{t-3}^2 \Bigr]
\biggr\} \cr
&\ +\ O(v^8).
&(4.2c) \cr
}$$
Expressions (4.2a), (4.2b), and (4.2c) account accurately for
backward excursions of length $2$, $4$, and $6$, respectively.

Note that the expansion of the ${\rm ln}[\hat \lambda_{k_{\perp}}
(t, \phi)]$ (the argument of ${\rm exp}\{ \}$ above) looks
rather like the expansion of $\hat \lambda_{k_{\perp}}(t, \phi)$
itself (eq. (3.11b)) except that coefficients have an additional
$1/n_1$ factor
$${\rm ln} \bigl[\hat \lambda_{k_{\perp}}(t, \phi) \bigr]
\ =\  \sum_{\{n_1, n_2,...\}} \Biggl[ {1 \over n_1} ~
\prod_{i=1}^{\infty}
{n_i+n_{i+1}-1 \choose n_{i+1}}
\Bigl(4v^2 C_{t-i}^2\Bigr)^{n_i} \biggl] ,
\eqno(4.3) $$
and, of course, the $n_1=0$ term is absent.
\ We can verify this expression with repeated use of the relation
$$\sum_{j=1}{1 \over j} {j+k-1 \choose k}x^j \ =\
\cases{ -{\rm ln}(1-x), &\ \ if $k=0$;\cr
{1 \over k}(1-x)^{-k} -{1 \over k}, & \ \ if $k \neq 0$,
\cr} \eqno(4.4)$$
which leads to
$${\rm ln} \bigl( \hat \lambda_{k_{\perp}} (t, \phi) \bigr)
\ =\ {\rm ln}\Bigl[D_{k_{\perp}}(t, \phi;2,s) \Bigr]
- {\rm ln}\Bigl[D_{k_{\perp}}(t, \phi;1,s)\Bigr]
\ +\ O(v^{2s+2}), \eqno(4.5)$$
which is simply the logarithm of eq. (3.3a).

The form given above is particularly convenient for calculating
$G^{b.e.}_{k_{\perp}}(0,T_o,\phi)$, the Green's function for
long walks that allow backward excursions, because
$G^{b.e.}_{k_{\perp}}(0,T_o,\phi)$ factors into
$G^{dir}_{k_{\perp}}(0,T_o,\phi)$ and a product of
$\hat \lambda$'s
$$
G^{b.e.}_{k_{\perp}}(0,T_o,\phi) \ =\
\prod_{t=1}^{T_o} ~\tilde \lambda_{k_{\perp}}(t, \phi)
\ =\ G^{dir}_{k_{\perp}}(0,T_o,\phi) \times
\prod_{t=1}^{T_o} ~\hat \lambda_{k_{\perp}}(t,\phi),
\eqno(4.6a) $$
and the exponential form converts that product into a sum:
$$\eqalignno{
G^{b.e.}_{k_{\perp}}(0,T_o,\phi) \ =\
& G^{dir}_{k_{\perp}}(0,T_o,\phi) \cr
&\times {\rm exp} \biggl\{ \sum_{t=1}^{T_o} \Bigl[
(2v)^2C_{t-1}^2 + (2v)^4\bigl[ {\textstyle{1 \over 2}}
C_{t-1}^4 + C_{t-1}^2C_{t-2}^2 \bigr]
+\ldots \Bigr] \biggr\}.
&(4.6b) \cr}$$

Let us return to the example considered in the previous section
(eqs. (3.5) and (3.6)) in which we have examined the effect of
short backward excursions (of length $2$ only) on the interference
pattern generated on the superlattice when the flux-per-plaquette
is commensurate $(\phi={\textstyle{p \over q}})$.
\ We are now in a position to extend that analysis.
\ Substituting $T_o=q$ and $\phi={\textstyle{p \over q}}$ into
eq. (4.6b), we find
$$\eqalignno{
G^{b.e.}_{k_{\perp}}(0,q,{\textstyle{p \over q }}) \ =\
& G^{dir}_{k_{\perp}}(0,q,{\textstyle{p \over q}}) \cr
&\times {\rm exp} \biggl\{
(2v)^2{\textstyle{q \over 2}}
+ (2v)^4\Bigl[ {\textstyle{7q \over 16}}
+ {\textstyle{q \over 8}} \cos \bigl(
{\textstyle{2 \pi p \over q}} \bigr) \Bigr]
+\ldots  \biggr\},
&(4.7) \cr}$$
where we have used the following relations for $\phi=
{\textstyle{p \over q}}$
$$\eqalignno{
\sum_{t=1}^q C_{t-1}^2 \ &=\ {q \over 2}
\ \ \ \ {\rm for} \ \ q\geq 2,
&(4.8a) \cr
\sum_{t=1}^q C_{t-1}^4 \ &=\ {3q \over 8}
\ \ \ \ {\rm for} \ \ q\geq 3,
&(4.8b) \cr
\sum_{t=1}^q C_{t-1}^2C_{t-2}^2 \ &=\ q\biggl[{1 \over 4}
+{1 \over 8} \cos \bigl( {\textstyle{2 \pi p \over q}}
\bigr) \biggr].
\ \ \ \ {\rm for} \ \ q\geq 3.
&(4.8c) \cr}$$
The exponential part represents the correction to the
directed-paths-only approach.
\ We see the same pattern here as before --- that on the
superlattice ($T_o=q$) the transverse momentum $k_{\perp}$ does
not enter the corrections ({\it i.e.} there are no structural
changes) at order $v^2$ or order $v^4$ provided $q \geq 3$,
suggesting that $k_{\perp}$ enters only at order $v^{2q}$.
\ Some of the sums needed to carry out this analysis to
higher order are:
$$\sum_{j=1}^{q} C_{t-j}^{2m} \ =\
{q \over 2^{2m}}{2m \choose m} \ \ \ \ {\rm if} \ \ q > m  ,
\eqno(4.9a)$$
$$\sum_{j=i}^{q} C_j^{2m}C_{j+1}^{2n} \ =\
{q \over 2^{2m+2n-1}} \Biggl[ \sum_{\ell=0}^n
{2m \choose m+\ell}{2n \choose n-\ell} \cos
\bigl({\textstyle {2 \pi p \ell \over q}} \bigr)
-{1 \over 2} {2m \choose m} {2n \choose n} \Biggr], \eqno(4.9b)$$
if $q>m+n$.
This last calculation is really a mixture of two special cases
--- small $v$ and commensurate flux ($\phi={\textstyle{p
\over q}}$).
\ We will now turn our attention to the commensurate case.

\vskip 0.1truein

\noindent{\bf iv.b. Commensurate flux}

In the case in which the flux-per-plaquette $\phi$ is a
rational multiple (${p \over q}$) of the flux quantum, the
terms in the continued fraction expression (eq. (3.2)) begin
to repeat themselves after $q$ terms.
\ For instance, for $\phi={1 \over 3}$, one has:
$$\hat \lambda_{k_{\perp}}(t,{\textstyle{1 \over 3}})
= {1 \over \displaystyle \ 1\ -\
{\strut 4v^2 C_{t-1}^2 \over \displaystyle \ 1\ -\
{\strut 4v^2 C_{t-2}^2 \over \displaystyle \ 1\ -\
\strut 4v^2 C_{t-3}^2 ~ \hat \lambda_{k_{\perp}}
(t,{\textstyle{1 \over 3}})}}}. \eqno(4.10a)$$
Simplifying the fraction yields:
$$\hat \lambda_{k_{\perp}}(t, {\textstyle{1 \over 3}})
\ =\ { 1 -4v^2C_{t-2}^2 - 4v^2 C_{t-3}^2 ~ \hat \lambda_{
k_{\perp}} (t,{\textstyle{1 \over 3}}) \over
1-4v^2C_{t-1}^2 -4v^2C_{t-2}^2 -4v^2C_{t-3}^2
{}~ \hat \lambda_{k_{\perp}} (t,{\textstyle{1 \over 3}}) }
. \eqno(4.10b)$$
Performing the same procedure for any $\phi={p \over q}$,
we find:
$$\hat \lambda_{k_{\perp}} (t,{\textstyle{p \over q}}
)\ =\ {D_{k_{\perp}}(t,{\textstyle{p \over q}};2,q-1)
-4v^2 C_{t-q}^2
D_{k_{\perp}}(t, {\textstyle{p \over q}}, 2,q-2)
{}~ \hat \lambda_{k_{\perp}}(t, {\textstyle{p \over q}})
\over D_{k_{\perp}}(t, {\textstyle{p \over q}};1,q-1)
-4v^2 C_{t-q}^2
D_{k_{\perp}}(t, {\textstyle{p \over q}}, 1,q-2)
{}~ \hat \lambda_{k_{\perp}}(t, {\textstyle{p \over q}}) } .
\eqno(4.11)$$

{}From here we conclude that $\hat \lambda_{k_{\perp}}
(t,{\textstyle{p \over q}})$ is the solution of the following
quadratic equation:
$${\cal A}_{k_{\perp}}(t, {\textstyle{p \over q}})
{}~ \Bigl[\hat \lambda_{k_{\perp}}(t, {\textstyle{p \over q}})
\Bigr]^2
\ +\ {\cal B}_{k_{\perp}}(t, {\textstyle{p \over q}})
{}~ \hat \lambda_{k_{\perp}}(t, {\textstyle{p \over q}})
\ +\ {\cal C}_{k_{\perp}}(t, {\textstyle{p \over q}})
\ =\ 0,\eqno(4.12a)$$
where
$$\eqalignno{
{\cal A}_{k_{\perp}}(t, {\textstyle{p \over q}})  \ =\
& 4v^2 C_{t-q}^2
D_{k_{\perp}}(t, {\textstyle{p \over q}};1,q-2),
&(4.12b)\cr
{\cal B}_{k_{\perp}}(t, {\textstyle{p \over q}})  \ =\
&- D_{k_{\perp}}(t, {\textstyle{p \over q}};1,q-1)
{}~-~4v^2C_{t-q}^2
 D_{k_{\perp}}(t, {\textstyle{p \over q}};2,q-2),
&(4.12c)\cr
{\cal C}_{k_{\perp}}(t, {\textstyle{p \over q}})  \ =\
& D_{k_{\perp}}(t, {\textstyle{p \over q}};2,q-1).
&(4.12d)\cr}$$
The solution of the quadratic equation is:
$$\hat \lambda_{k_{\perp}}(t, {\textstyle{p \over q}})
\ =\ { ~-~ {\cal B}_{k_{\perp}}(t, {\textstyle{p \over q}})
{}~-~ \sqrt{ ~{\cal B}^2_{k_{\perp}}(t, {\textstyle{ p \over q}})
{}~-~4~ {\cal A}_{k_{\perp}}(t, {\textstyle{p \over q}})
{}~ {\cal C}_{k_{\perp}}(t, {\textstyle{p \over q}}) } \over
2 ~{\cal A}_{k_{\perp}}(t, {\textstyle{p \over q}}) } .
\eqno(4.13) $$

We have used this last expression to calculate $\hat
\lambda_{k_{\perp}}(t, {\textstyle{p \over q}})$ for
relatively small $q$.
\ Some algebra and trigonometry  reveal that:
$$\eqalignno{
\hat \lambda_{k_{\perp}}(t,1) \ =\
& { 1- \sqrt{ 1-16v^2C_{t-1}^2 } \over 8v^2C_{t-1}^2 } ,
&(4.14a) \cr
\hat \lambda_{k_{\perp}}(t,{\textstyle{1 \over 2}}) \ =\
& { 1- 4v^2 + 8v^2C_{t-2}^2
- \sqrt{ (1-4v^2)^2-16v^4\sin^2 (2k_{\perp}) } \over
8v^2C_{t-2}^2 } ,
&(4.14b) \cr
\hat \lambda_{k_{\perp}}(t,{\textstyle{1 \over 3}}) \ =\
& { 1- 6v^2 + 8v^2C_{t-3}^2
- \sqrt{ (1-6v^2)^2-16v^6\cos^2 (3k_{\perp}) }
\over 8v^2C_{t-3}^2 (1-4v^2C_{t-1}^2)}.
&(4.14c) \cr }$$
One can see a pattern arising from these expressions.
\ For instance, we can see here and show in general that
the discriminant $({\cal B}^2-4 {\cal A} {\cal C})$ is
independent of $t$.

Note that $q=1$ corresponds to one flux quantum per plaquette or
equivalently no field at all.
\ The small field limit, on the other hand, would require
taking $q$ to infinity.
\ Fortunately, this limit is accessible by other means.

\vskip 0.1truein

\noindent{\bf iv.c. The small field limit}

The third special case we consider here is when the magnetic
field is small.
\ Toward this end, let us perform a Taylor expansion of $\hat
\lambda_{k_{\perp}}(t,\phi)$ around $(k_{\perp}-t\phi\pi)$:
$$\hat \lambda_{k_{\perp}}(t,\phi) \ =\
\sum_{i=0}^{\infty} \Lambda_{k_{\perp}}^{(i)}(t,\phi)
{}~\phi^i. \eqno(4.15)$$
It is our goal in this section to calculate the $i=0$, $1$
and $2$ terms $\Lambda^{(i)}_{k_{\perp}}(t,\phi)$ above.

Instead of expanding $\hat \lambda_{k_{\perp}}(t,\phi)$
directly, we will begin by expanding $\hat
\lambda^{(j)}_{k_{\perp}}(t,\phi)$.
\ Recalling eq. (3.10a) for $\hat \lambda^{(j)}_{k_{\perp}}
(t,\phi)$, we clearly need an expansion of $C_{t-j}^{\ell}$
around $(k_{\perp}-t\phi\pi)$:
$$C_{t-j}^{\ell} \ =\ C_t^{\ell} ~-~{j\ell\phi \pi}
C_t^{\ell-1}S_t ~+~{j^2 \phi^2 \pi^2 \over 2} \Bigl[ \ell(\ell-1)
C_t^{\ell-2}S_t^2 - \ell C_t^{\ell} \Bigr] ~+~
..., \eqno(4.16)$$
where $S_t= {\rm sin}(k_{\perp}-t\phi \pi)$.
\ Below we present the expansions of $\hat
\lambda_{k_{\perp}}^{(j)}(t, \phi)$ to order $\phi^2$ for
$j=0,1,\ldots,6$:
$$\eqalignno{
\hat \lambda^{(0)}_{k_{\perp}}(t, \phi) = & 1
&(4.17a)\cr
\hat \lambda^{(1)}_{k_{\perp}}(t, \phi) = &(2v)^2 \Bigl[
C_t^2 - 2 \pi\phi C_tS_t + {\pi^2 \phi^2 \over 2}
(2S_t^2-2C_t^2)-...\Bigr]
&(4.17b)\cr
\hat \lambda^{(2)}_{k_{\perp}}(t, \phi) = &(2v)^4 \Bigl[
2C_t^4 - 10 \pi \phi C_t^3S_t + {\pi^2 \phi^2 \over 2}
(38C_t^2S_t^2-14C_t^4)
-...\Bigr]
&(4.17c)\cr
\hat \lambda^{(3)}_{k_{\perp}}(t, \phi) = &(2v)^6 \Bigl[
5C_t^6 - 44 \pi \phi C_t^5S_t + {\pi^2 \phi^2 \over 2}
(332C_t^4S_t^2
-76C_t^6)-...\Bigr]
&(4.17d)\cr
\hat \lambda^{(4)}_{k_{\perp}}(t, \phi) = &(2v)^8 \Bigl[
14C_t^8 - 186 \pi \phi C_t^7S_t + {\pi^2 \phi^2 \over 2}
(2246C_t^6S_t^2-374C_t^8)-...\Bigr]
&(4.17e)\cr
\hat \lambda^{(5)}_{k_{\perp}}(t, \phi) = &(2v)^{10} \Bigl[
42C_t^{10} - 772 \pi \phi C_t^9S_t + {\pi^2 \phi^2 \over 2}
(13348C_t^8S_t^2-1748C_t^{10})-...\Bigr]
&(4.17f)\cr
\hat \lambda^{(6)}_{k_{\perp}}(t, \phi) = &(2v)^{12} \Bigl[
132C_t^{12} - 3172 \pi \phi C_t^{11}S_t \cr
&\ \ \ \ \ \ \ \ \ \ \ \ \ \ \
+ {\pi^2 \phi^2 \over 2}
(73340C_t^{10}S_t^2-7916C_t^{12})-...\Bigr]
&(4.17g)\cr} $$

Next, we collect terms of the same order in $\phi$ and
resum those series.
\ Toward that end, we define:
$$\eqalignno{
\Lambda_{k_{\perp}}^{(0)} \ =\ &\sum_{j=0}^{\infty}
\ N_j \ (2vC_t)^{2j}, &(4.18a) \cr
\Lambda_{k_{\perp}}^{(1)} \ =\ &-{2 \pi S_t \over C_t}
\sum_{j=0}^{\infty}
\ P_j \ (2vC_t)^{2j}, &(4.18b) \cr
\Lambda_{k_{\perp}}^{(2)} \ =\ &
{\pi^2 S_t^2 \over 2 C_t^2}
\sum_{j=0}^{\infty} \ Q_j \ (2vC_t)^{2j} \ -\
{\pi^2 \over 2}
\sum_{j=0}^{\infty} \ R_j \ (2vC_t)^{2j}. &(4.18c) \cr}$$
{}From eqs. (4.17a-g), we can conclude, for instance, that
$$\bigl\{P_j \bigr\} \ =\ \bigr\{~0,\ 1,\ 5,\ 22,\ 93,\
386,\ 1586, \ \ldots~\bigr\}. \eqno(4.19)$$

These four series are summed as follows:
$$\eqalignno{
\sum_{j=0}N_j~x^j \ =\ &{1 \over 2x} \Bigl[
1-(1-4x)^{1/2}\Bigr],
&(4.20a)\cr
\sum_{j=0}P_j~x^j \ =\ &{1 \over 2} \Bigl[
(1-4x)^{-1}-(1-4x)^{-1/2}\Bigr],
&(4.20b)\cr
\sum_{j=0}Q_j~x^j \ =\ &{5 \over 2}(1-4x)^{-5/2}
-2(1-4x)^{-2}-2(1-4x)^{-3/2} \cr
&\ \ \ \ +(1-4x)^{-1} +{1 \over 2}(1-4x)^{-1/2},
&(4.20c)\cr
\sum_{j=0}R_j~x^j \ =\ &(1-4x)^{-3/2} - (1-4x)^{-1}.
&(4.20d) \cr}$$

Using these we arrive at:
$$\Lambda_{k_{\perp}}^{(0)}(t,\phi) \ =\
{1- \sqrt{1-16v^2C_t^2} \over 8v^2C_t^2 }, \eqno(4.21a)$$
$$\Lambda_{k_{\perp}}^{(1)}(t,\phi) \ =\
-~{8 \pi v^2 C_tS_t \over 1 -16 v^2 C_t^2}
\ \Lambda_{k_{\perp}}^{(0)}(t,\phi) \eqno(4.21b)$$
and
$$\eqalignno{
\Lambda^{(2)}_{k_{\perp}}(t,\phi) \ =\
2\pi^2v^2 U^{-3}\Biggl\{ &2(S_t^2-C_t^2) \cr
&~+~ S_t^2 U^{-2}
\Bigl[ 5+U\Bigr]~\Bigl[1-U^2\Bigr]
\Biggr\}\Lambda^{(0)}_{k_{\perp}}, &(4.21c) \cr}$$
where $U=(1-16 v^2 C_t^2)^{-1/2}$.

Note that the linear term breaks the $\phi \rightarrow -
\phi$ symmetry.
\ Two comments should be made: \ (i) $\phi \rightarrow -
\phi$ with $x \rightarrow -x$ (or equivalently $k_{\perp}
\rightarrow -k_{\perp}$) is still a symmetry and (ii)
most importantly all physical properties which depend on the
absolute value $|G|$ will be invariant under $\phi \rightarrow
- \phi$ or $x \rightarrow -x$ separately.
\ This is consistent with the expectation that no Lorentz force
(and no parity symmetry breaking) should occur for states which
do not carry current density.
\ Such effects will be seen only if the dependence of $G$ in the
$t$ direction goes like ${\rm e}^{ik_{\parallel}t}$ so that
the states have a nonvanishing current density in the $t$
direction. [14]

\vskip 10pt

\noindent {\bf V. The Full Green's Function}

\vskip 5pt

In this section, we show how to obtain the full Green's function
${\cal G}_{k_{\perp}}(0,T_o,\phi)$, which is the appropriately
weighted sum of all walks connecting columns $0$ and $T_o$.
\ (Recall that the $2D$ problem was reduced to $1D$.)
\ Thus far, we have calculated $\tilde \lambda_{k_{\perp}}
(t,\phi)=G^{b.e.}_{k_{\perp}}(t-1,t,\phi)$,
which includes walks which advance from column $t-1$ to column $t$
with any amount of backward excursion prior to the forward step.
\ In this section we will find it more convenient to use the
$G^{b.e.}_{k_{\perp}}(t-1,t,\phi)$ notation.
\ We have also already seen the product:
$$G^{b.e.}_{k_{\perp}}(0,T_o,\phi) \ =\
\prod_{t=0}^{T_o-1} G^{b.e.}_{k_{\perp}}(t,t+1;\phi); \eqno(5.1)$$
which supplies the Green's function for the restricted walks that begin
at the origin and end in the $T_o$ column, again with the last step
being the only one connecting the $T_o-1$ column to the $T_o$ column.
\ This is the Green's function for the first time the walker
lands in column $T_o$.

Now we must allow the walker to go beyond $T_o$ or to turn
back and return later to $T_o$.
\ Let us denote this sum by ${\cal G}^{full}_{k_{\perp}}
(T_o,T_o; \phi)$; it is
the Green's function for all walks starting in column $T_o$ and
returning to that column.
\ Note that ${\cal G}^{full}_{k_{\perp}}(T_o,T_o;\phi)$ obeys the
following recursion relation:

\vfill
\eject

$$\eqalignno{
{\cal G}^{full}_{k_{\perp}}(T_o,T_o;\phi) \ =\ 1 \ +\ &
4 v^2 C_{T_o}^2
{}~G^{b.e.}_{k_{\perp}}(T_o-1,T_o-1,\phi)
{}~{\cal G}^{full}_{k_{\perp}}(T_o,T_o,\phi) \cr
+ \ & 4 v^2 C_{T_o+1}^2
{}~G^{f.e.}_{k_{\perp}}(T_o+1,T_o+1\phi)
{}~{\cal G}^{full}_{k_{\perp}}(T_o,T_o;\phi), &(5.2a) \cr}$$
where $G^{f.e.}$ is the same Green's function as we have
already found except that the initial position is to the right of the
endpoint.
\ Therefore, the formal solution is:
$${\cal G}^{full}_{k_{\perp}}(T_o,T_o; \phi) =\Bigl[ 1 -
4v^2C_{T_o}^2 G^{b.e.}_{k_{\perp}}(T_o-1,T_o-1,\phi)
-4v^2C_{T_o+1}^2 G^{f.e.}_{k_{\perp}}(T_o+1,T_o+1;\phi)
\Bigr]^{-1}. \eqno(5.2b)$$

Finally, the full Green's function
${\cal G}^{full}_{k_{\perp}}(0,T_o;\phi)$
is simply the product:
$${\cal G}^{full}_{k_{\perp}}(0,T_o; \phi) \ =\
G^{b.e.}_{k_{\perp}}(0,T_o;\phi )
\ {\cal G}^{full}_{k_{\perp}}(T_o,T_o;\phi ),
\eqno(5.3a)$$
$${\cal G}^{full}_{k_{\perp}}(0,T_o,\phi) ={
{\displaystyle\prod_{t=0}^{T_o-1} G^{b.e.}_{k_{\perp}}
(t,t+1;\phi)} \over \Bigl[ 1 -  4v^2C_{T_o}^2
G^{b.e.}_{k_{\perp}}(T_o-1,T_o-1;\phi)
-  4v^2C_{T_o+1}^2 G^{f.e.}_{k_{\perp}}(T_o+1,T_o+1;\phi) \Bigr]}.
\eqno(5.3b)$$

This last equation is the central equation of this paper: \
It gives the formal expression for the Green's function of an
electron on a lattice in a magnetic field.
\ The dependence on $x$ can be obtained by an inverse Fourier
transform with respect to ${\rm exp}(ik_{\perp}x)$.
\ In principle, all physical properties may be obtained from
this expression.
\ This is certainly true for $E$ outside the spectrum.
\ For the energies within the \lq \lq butterfly," one may still
use this formula for walks which do not reach the boundaries of
the system (or wrap around the torus for if periodic boundary
conditions are assumed, other delicate questions then arise [13]
if the two sizes $L_{\perp}$ and $L_{\parallel}$ are not integer
multiples of $q$.)

In the next section, we look at the Green's function for charged
particles propagating within the staggered-flux configurations
that have been studied in the context of high-$T_c$ superconductivity.

\vskip 10pt

\noindent {\bf VI. Staggered Flux Configurations}

\vskip 5pt

Beginning with the work of Affleck and Marston [11], the notion of
staggered flux configurations was introduced and studied in the
context of theories of high-$T_c$ superconductivity.
\ It may, therefore, be useful to have expressions for electron
propagation in such flux configurations.
\ On a square lattice, such a configuration will be given by a
chessboard arrangement of interpenetrating sets of plaquettes
with (say) $+\phi$ on the white squares and $-\phi$ on the
black ones.
\ (See Figure 2.)
\ The most studied case is $\phi=\pi$, for which the phase factors
acquired around the plaquettes are $\pm1$, so that time reversal
symmetry is unbroken.
\ The sum of walks in this particular case was obtained by
Khveshchenko, Kogan and Nechaev [15].
\ In the present section, we obtain these sums for arbitrary
staggered flux $\pm \phi$.

First, we must choose a gauge:
\ this time we assign the phases only to the vertical bonds of a
square lattice.
\ Considering them as directed upwards, we assign a factor
$\gamma={\rm e}^{i\phi/2}$ for bonds pointing from sublattice
$A$ to sublattice $B$ (shown as dashed in Fig. 2) and
$\gamma^*={\rm e}^{-i\phi/2}$ to the bonds pointing from
sublattice $B$ to $A$ (solid in Fig. 2).
\ So a staggered arrangement of phases $\pm \phi/2$ is assigned
to the (upward-pointing) vertical bonds.

Next, we assign a factor $\alpha$ ($\alpha^{-1}$) to a step in
the positive (negative) $x$ direction and similarly $\beta$
($\beta^{-1}$) for the $y$ direction.
\ A natural choice is $\alpha = {\rm e}^{ik_x}$ and $\beta =
{\rm e}^{ik_y}$ which yields the Fourier transform of the generating
functions.
\ We define the following generating functions:
\ first, the local generating functions for a step
$$\eqalignno{
A \rightarrow B: \ \ \ & g_A \ =\ \alpha + \alpha^{-1} +
\gamma (\beta + \beta^{-1}) \cr
B \rightarrow A: \ \ \ & g_B \ =\ \alpha + \alpha^{-1} +
\gamma^{-1} (\beta + \beta^{-1}), &(6.1)\cr }$$
one for each sublattice; then a global generating function
for all paths of length $2L$:
$$G^{stag}_{2L}(\phi) \ =\ g_A^Lg_B^L \ =\ \Bigl[ 4(\cos^2 k_x
+\cos^2 k_y + 2 \cos {\phi \over 2} \cos k_x \cos k_y )
\Bigr]^L. \eqno(6.2)$$
The probability amplitude to reach a given site $(x,y)$ is the
coefficient of $\alpha^x \beta^y$ (up to a trivial normalization)
or the inverse Fourier transform:
$${\cal P}^{stag}_{2l}(\phi) \ \propto \
{1 \over (2 \pi)^2}
\int_{-\pi}^{\pi} \int_{-\pi}^{\pi} dk_x \ dk_y \ G^{stag}_{2L}
(\phi) \ {\rm e}^{-ik_xx-ik_yy}, \eqno(6.3)$$
where the system size is assumed to be infinite and we have thus
gone over to integrals over the momenta.
\ Note that in order to obtain the normalized Green's function, $g_A$
and $g_B$ must each be multiplied by ${1 \over 2}$ and $G_{2L}$
by $2^{-2L}$.

Of particular interest are the expressions for walks which
return to their initial point.
\ The weighted closed magnetic walks are given by the coefficient
of $\alpha^0\beta^0=1$ in the product:
$$G^{stag}_{2L}(\phi) \ =\ \Bigl[ (\alpha + \alpha^{-1}) +
\gamma (\beta + \beta^{-1}) \Bigl]^L
\Bigl[ (\alpha + \alpha^{-1}) +
\gamma^{-1} (\beta + \beta^{-1}) \Bigl]^L,\eqno(6.4)$$
where $\gamma={\rm e}^{i \phi/2}$.
\ Toward this end, we expand the above products as follows:
$$\eqalignno{
\Bigl[ (\alpha + \alpha^{-1}) +
\gamma (\beta+\beta^{-1}) \Bigr]^L \ =\ &
\sum_{L_x^{(A)}} {L \choose L_x^{(A)} }
(\alpha + \alpha^{-1})^{L_x^{(A)}}
\Bigl[ \gamma (\beta + \beta^{-1})\Bigr]^{L_y^{(A)}} &(6.5a) \cr
\Bigl[ (\alpha + \alpha^{-1}) +
\gamma^{-1}(\beta + \beta^{-1}) \Bigr]^L \ =\ &
\sum_{L_x^{(B)}} {L \choose L_x^{(B)} }
(\alpha + \alpha^{-1})^{L_x^{(B)}}
\Bigl[ \gamma^{-1} (\beta + \beta^{-1})\Bigr]^{L_y^{(B)}}
&(6.5b) \cr}$$
where $L_x^{(A)}$ ($L_y^{(A)}$) is the number of horizontal
(vertical) steps initiated from the $A$ sublattice and likewise
for the $B$ sublattice.
\ Consequently, $L_x^{(A)}+L_y^{(A)}=L_x^{(B)}+L_y^{(B)}=L$.
\ Next, we expand the remaining $\alpha$ products, introducing the
variables $\ell_{+x}^{(A)}$ and $\ell_{+x}^{(B)}$ which represent
the number of steps in the positive $x$ direction initiated
on the $A$ and $B$ sublattices, respectively.
\ The coefficients of $\alpha$ will thus be:
$$\sum_{\ell_{+x}^{(A)}}^{L_x^{(A)}}
\sum_{\ell_{+x}^{(B)}}^{L_x^{(B)}}
{L_x^{(A)} \choose \ell_{+x}^{(A)}}
\alpha^{2\ell_{+x}^{(A)} -L_x^{(A)}}
{L_x^{(B)} \choose \ell_{+x}^{(B)}}
\alpha^{2\ell_{+x}^{(B)} -L_x^{(B)}}.
\eqno(6.6)$$
To obtain the $\alpha^0$ term, the following relation must hold:
$$2(\ell_{+x}^{(A)} + l_{+x}^{(B)}) \ =\ L_x^{(A)} + L_x^{(B)} \
=\ L_x, \eqno(6.7)$$
where $L_x$ is, of course, the total number of horizontal steps.
\ Note that $L_x$ must be even.
\ Replacing $\ell_{+x}^{(B)}$ by ${L_x \over 2} -\ell_{+x}^{(A)}$,
we find that the coefficient of $\alpha^0$ is:
$$\sum_{l_{+x}^{(A)}}^{L_x^{(A)}} {L_x^{(A)} \choose \ell_{+x}^{(A)}}
{L_x^{(B)} \choose {L_x \over 2}-\ell_{+x}^{(A)}} \ =\
{L_x \choose {L_x \over 2}}. \eqno(6.8)$$
Repeating the same procedure for the $y$ direction and combining
the results yields:
$$\eqalignno{
G^{stag}_{2L}(\phi) \ =\ &
\sum_{L_x^{(A)}} \sum_{L_x^{(B)}}
{L \choose L_x^{(A)}} {L_x \choose {L_x \over 2}}
{L \choose L_x^{(B)}} {L_y \choose {L_y \over 2}}
\gamma^{L_y^{(A)}-L_y^{(B)}} &(6.9a) \cr
&\sum_{L_x=0}^L \sum_{L_x^{(A)}=0}^{L_x}
{L \choose L_x^{(A)}}
{L_x \choose {L_x \over 2}}
{L \choose L_x - L_x^{(A)}}
{L_y \choose {L_y \over 2}}
\gamma^{2L_y^{(A)}-L_y} &(6.9b) \cr
= \ &\sum_{L_x}
{L_x \choose {L_x \over 2}}
{2L-L_x \choose L-{L_x \over 2}}
\sum_{L_x^{(A)}}
{L \choose L_x^{(A)} }
{L \choose L_x - L_x^{(A)}}
\gamma^{L_x-2L_x^{(A)}},&(6.9c) \cr}$$
where we have used $L_y^{(A)}=L-L_x^{(A)}$ and $L_y = 2L
-L_x$.

Using eq. (6.9) we have found $G^{stag}_{2L}$ for several
$L$'s.
\ They are:
$$\eqalignno{
G^{stag}_0 \ =\ &1 &(6.10a)\cr
G^{stag}_2 \ =\ &4 &(6.10b)\cr
G^{stag}_4 \ =\ &28 + 8 \cos \phi &(6.10c)\cr
G^{stag}_6 \ =\ &256 + 144 \cos \phi &(6.10d) \cr
G^{stag}_8 \ =\ &2716 + 2112 \cos \phi + 72 \cos 2 \phi
&(6.10e)\cr
G^{stag}_{10} \ =\ &31504 + 29600 \cos \phi + 2400 \cos 2 \phi
&(6.10f)\cr
G^{stag}_{12} \ =\ &387136 + 411840 \cos \phi + 54000 \cos 2 \phi
+ 800 \cos 3 \phi &(6.10g)\cr
G^{stag}_{14} \ =\ &4951552 + 5752992 \cos \phi + 1034880 \cos 2 \phi
+ 39200 \cos 3 \phi &(6.10h)\cr
G^{stag}_{16} \ =\ &65218204 + 80950016 \cos \phi + 18267200 \cos 2 \phi
+ 1191680 \cos 3 \phi \cr
&+ 9800 \cos 4 \phi &(6.10i)\cr
G^{stag}_{18} \ =\ &878536624 + 1148084928 \cos \phi + 307577088 \cos 2 \phi
+ 29070720 \cos 3 \phi \cr
&+ 635040 \cos 4 \phi &(6.10j)\cr
G^{stag}_{20} \ =\ &12046924528 + 16407496800 \cos \phi + 5030575200 \cos
2\phi + 625312800 \cos 3 \phi \cr
&+ 24343200 \cos 4 \phi + 127008 \cos 5 \phi
&(6.10k)\cr}$$
$G^{stag}_{2L}$ can also be obtained as the coefficient of
$v^{2L}$ in the following expression:
$$\eqalignno{
{1 \over 2 \pi} \int_0^{2 \pi} dk_{\perp}
 \Bigl\{ 1 &-8v^2 -
8v^2\cos (\phi /2) \cos (2k_{\perp})
+16v^4\sin^2(\phi /2) \sin^2(2k_{\perp}) \Bigr\}^{-{1 \over
2}}. &(6.10 \ell) \cr}$$

As we work toward procuring a more general expression,
let us concentrate on the following sum:
$$\sum_{L_x^{(A)}}^{L_x}
{L \choose L_x^{(A)}}
{L \choose L_x-L_x^{(A)}}
\gamma^{L_x-2L_X^{(A)}}, \eqno(6.11a)$$
which is the latter portion of eq. (6.9c).
\ Introducing $L_x^{(B)}=L_x-L_x^{(A)}$ this may be written as:
$$\eqalignno{
\sum_{L_x^{(A)}}^L \sum_{L_x^{(B)}}^L
& {L \choose L_x^{(A)}}
{L \choose L_x^{(B)}}
\gamma^{L_x^{(B)}-L_x^{(A)}}
\delta \Bigl[ L_x-L_x^{(A)}-L_x^{(B)} \Bigr] \cr
&={1 \over 2 \pi} \int_0^{2 \pi} d \theta \sum_{L_x^{(A)}}
\sum_{L_x^{(B)}}
{L \choose L_x^{(A)}}
{L \choose L_x^{(B)}}
\gamma^{L_x^{(B)}-L_X^{(A)}}
{\rm e}^{-i\theta [L_x-L_x^{(A)}-L_x^{(B)}]} \cr
&={1 \over 2 \pi} \int_0^{2 \pi} d \theta \
\Bigl[1+\gamma {\rm e}^{i \theta} \Bigr]^L
\Bigl[1+\gamma^{-1} {\rm e}^{i \theta} \Bigr]^L
{\rm e}^{-i \theta L_x}  \cr
&={1 \over 2 \pi} \int_0^{2 \pi} d \theta \
{\rm e}^{i \theta (L-L_x)}
\Bigl[ 2 \cos \theta + (\gamma + \gamma^{-1}) \Bigr]^L \cr
&={2^L \over 2 \pi} \int_0^{2 \pi} d \theta \
{\rm e}^{i \theta (L-L_x)}
\Bigl[ \cos \theta + \cos {\phi \over 2} \Bigr]^L
&(6.11b) \cr }$$

We may now use the identity:
$${1 \over 2 \pi} \int_0^{2 \pi} d \theta
\Bigl[ \mu + \sqrt{\mu^2-1} \cos \theta \Bigr]^n
{\cos m\theta \choose \sin m\theta } \ =\
{n! \over (n+m)!} {P_n^m(\mu) \choose 0} \eqno(6.12)$$
We identify $\cos { \phi \over 2} = \mu / \sqrt{\mu^2-1}$ which
implies $\mu=\pm i \cot {\phi \over 2}$
such that the integral becomes:
$${2^L \over 2 \pi} {1 \over (\mu^2 -1)^{L \over 2}}
\int_0^{2 \pi} d \theta \ {\rm e}^{i \theta (L-L_x)}
\Bigl[ \mu + \sqrt{\mu^2-1}\cos \theta \Bigr]^L. \eqno(6.13)$$
Therefore for the sum
$$\sum_{L_x^{(A)}}^{L_x}
{L \choose L_x^{(A)}}
{L \choose L_x-L_x^{(A)}}
\gamma^{L_x-2L_x^{(A)}}\
= \ (-i)^{L \over 2} (\sin {\phi \over 2})^{L \over 2}
{2^L L! \over (2L-L_x)! }
P_L^{(L-L_x)}(i \cot {\phi \over 2}) \eqno(6.14)$$

The final result is:
$$G^{stag}_{2L} \ =\ C(L) \sum_{L_x=0}^{2L}
{L_x \choose {L_x \over 2}}
{2L-L_x \choose L-{L_x \over 2}}
{1 \over (2L-L_x)!}
P_L^{(L-L_x)}(i \cot {\phi \over 2}) \eqno(6.15a)$$
with
$$C(L) \ =\ (-i)^{L \over 2} \bigl( \sin {\phi \over 2}
\bigr)^{L \over 2} 2^L L! \eqno(6.15b)$$
\ We find agreement with the known cases [15] of $\phi=0$
for which eq. (6.14) yields ${2L \choose L_x}$ and
$\phi=\pi$ for which it gives ${L \choose {L_x \over 2}}$.

One can generalize the expressions to walks that do not
close.
\ To reach a given point ($x_o$,$y_o$) the only difference will be
in the summation over $l_{+x}^{(A)}$ since what we need now is
the coefficient of $\alpha^{x_o}$ or
$$2 \bigl( l_{+x}^{(A)}+l_{+x}^{(B)} \bigr) -
\bigl( L_x^{(A)} + L_x^{(B)} \bigr) \ =\ x_o. \eqno(6.16)$$
Therefore $l_{+x}^{(B)}={L_x \over 2}+{x_o \over 2} - l_{+x}^{(A)}$
and the sum is:
$$\sum_{l_{+x}^{(A)}=0}^{L_x^{(A)}}
{L_x^{(A)} \choose l_{+x}^{(A)}}
{L_x^{(B)} \choose {L_x \over 2} + {x_o \over 2} -l_{+x}^{(A)}}
\ =\ {L_x \choose {L_x \over 2} + {x_o \over 2}}. \eqno(6.17)$$
Likewise (to get the coefficient of $\beta^{y_o}$) the
summation in the $y$ direction yields:
$$\gamma^{L_y^{(A)}-L_y^{(B)}}
{L_y \choose {L_y \over 2}+{y_o \over 2}}, \eqno(6.18)$$
and the final expression changes to:
$$G^{stag}_{2L} \ =\
C(L)\sum_{L_x=0}^{2L}
{L_x \choose {L_x \over 2} +{x_o \over 2}}
{2L-L_x \choose L -{L_x \over 2} +{y_o \over 2}}
{1 \over (2L-L_x)!}
P_L^{(L-L_x)} (i \cot {\phi \over 2}). \eqno(6.19) $$
Alternatively, one could choose $\alpha={\rm e}^{ik_x}$ and
$\beta={\rm e}^{ik_y}$ and study $G_{2L}(\alpha, \beta)$ which
can then be Fourier transformed back into real space.

\vskip 10pt

\noindent {\bf VII. Discussion}

\vskip 5pt

In this paper we have derived expressions for the square-lattice
Green's function of a charged particle in a magnetic field.
\ We have also obtained similar expressions for the so-called
staggered flux configurations with arbitrary flux $\pm \phi$.
\ The results were obtained within the Peierls ansatz in which
the magnetic field is represented by an extra phase acquired by
the nearest-neighbor hopping term.

Our starting point was the directed-paths-only approach initiated
in the study of strongly localized wave functions.
\ From there, we have shown how to include systematically the
returning loops and obtain the Green's function for walks with
backward excursions, and then we have demonstrated how the full
Green's functions were realized in terms of the former.

Our main goal was to see if the inclusion of small returning loops
would modify in some essential way the behavior found in the
directed-paths approach.
\ The analysis of eqs. (3.5), (3.6) and (4.1) through
(4.8) shows that the answer is negative in the
sense that the product over cosines with changing phases obtained
for directed paths is modified into a product of more complicated
trigonometric expressions containing the same cosines.
\ Of course, the behavior will change but the essential features: \
strong sensitivity to the value of the field, simple scaling in $q$
for the commensurate case $\phi=p/q$, and an aperiodic but deterministic
behavior for irrational $\phi$, all are going to survive the addition
of the small returning loops.
\ Repeating the analysis done in ref. 10. for the product of the
new trigonometric expressions would be a formidable challenge,
since even for the former case, the analysis was a substantial and
sophisticated task.

Our expressions might aid in the study of other interesting issues
as well:
\ For example, the transmission through a slab of finite width in
which the wave function does not decay all the way to zero, might
also be studied.
\ An especially interesting question concerns the emergence of a
Lorentz force (and the associated parity symmetry breaking) in the
case which deals with such a current carrying state.

Another interesting issue is how the continuum limit may be
approached:
\ Naively, one would expect that if the magnetic flux is small
such that the magnetic length is much longer than the lattice
spacing (but smaller than the system size) the lattice results will
coincide with those obtained in the continuum.
\ In particular, the exponential decay should change into a
Gaussian $\sim {\rm exp}\{-cBr^2\}$ behavior. [4]

Finally, these Green's functions should be correct (analytically)
for energies within the band.
\ One should, of course, add a small imaginary part to the energy
(and define advanced and retarded Green's functions) to deal with
the singular behavior at the eigenenergies.
\ In addition, the question of boundary conditions for walks
hitting the boundaries should be handled with special care.
\ In principle at least, all physical information could be extracted
from these Green's functions.
\ It will certainly not be straightforward to extract physically
relevant information from the formal expressions, but in view of the
importance of such information for many physical problems of
interest (especially the many-body extensions studied in the context
of the Fractional Quantum Hall Effect, high-$T_c$ superconductivity,
anyonic physics, etc.), it may be worthwhile to pursue those
directions.

\vskip 0.5truein

\noindent {\bf Acknowledgments}

Acknowledgment is made to the donors of The Petroleum Research
Fund, administered by the ACS, for support of this research.

\vfill
\eject

{\centerline {\bf Appendix} }

In section 3 we have introduced the function $\tilde
\lambda_{k_{\perp}}(t, \phi)$, the eigenvalue
of the transfer matrix $\tilde T(t,\phi)$ which
allows any amount of ``backward excursion'' followed
by a single forward step.
\ We called $\tilde \lambda_{k_{\perp}}^{(j)}(t,
\phi)$ the contribution with path length $(2j+1)$.
\ Using the recursion relation (3.9), we have
generated $\tilde \lambda_{k_{\perp}}^{(j)}(t,
\phi)$ for $j=0,1,\ldots,5$.
\ They are:
$$\eqalignno{
\tilde \lambda^{(0)}_{k_{\perp}}(t, \phi) \ =\ &(2v)C_t
&(AI.1a)\cr
\tilde \lambda^{(1)}_{k_{\perp}}(t, \phi) \ =\ &(2v)^3C_t
C_{t-1}^2
&(AI.1b)\cr
\tilde \lambda^{(2)}_{k_{\perp}}(t, \phi) \ =\ &(2v)^5C_t \Bigl[
C_{t-1}^4 + C_{t-1}^2C_{t-2}^2 \Bigr]
&(AI.1c)\cr
\tilde \lambda^{(3)}_{k_{\perp}}(t, \phi) \ =\ &(2v)^7C_t \Bigl[
C_{t-1}^6 + 2C_{t-1}^4C_2^2+ C_{t-1}^2C_{t-2}^4 + C_{t-1}^2
C_{t-2}^2C_{t-3}^2\Bigr]
&(AI.1d)\cr
\tilde \lambda^{(4)}_{k_{\perp}}(x, \phi) \ =\ &(2v)^9C_t\Bigl[
C_{t-1}^8 + 3C_{t-1}^6C_{t-2}^2 + 3 C_{t-1}^4C_{t-2}^4
+2 C_{t-1}^4C_{t-2}^2C_{t-3}^2 \cr
&\ \ \ \ \ \ \ \ \ \ \
+ C_{t-1}^2C_{t-2}^6 +2C_{t-1}^2C_{t-2}^4C_{t-3}^2
+ C_{t-1}^2C_{t-2}^2C_{t-3}^4 \cr
&\ \ \ \ \ \ \ \ \ \ \
+ C_{t-1}^2C_{t-2}^2C_{t-3}^2C_{t-4}^2 \Bigr]
&(AI.1e)\cr
\tilde \lambda^{(5)}_{k_{\perp}}(t, \phi) \ =\ &(2v)^{11}C_t\Bigl[
C_{t-1}^{10} + 4C_{t-1}^8C_{t-2}^2 + 6C_{t-1}^6C_{t-2}^4
+3 C_{t-1}^6C_{t-2}^2C_{t-3}^2 \cr
&\ \ \ \ \ \ \ \ \ \ \
+ 4C_{t-1}^4C_{t-2}^6 +6C_{t-1}^4C_{t-2}^4C_{t-3}^2
+ 2C_{t-1}^4C_{t-2}^2C_{t-3}^4 \cr
&\ \ \ \ \ \ \ \ \ \ \
+ 2C_{t-1}^4C_{t-2}^2C_{t-3}^2C_{t-4}^2
+ C_{t-1}^2C_{t-2}^8
+3C_{t-1}^2C_{t-2}^6C_{t-3}^2 \cr
&\ \ \ \ \ \ \ \ \ \ \
+ 3C_{t-1}^2C_{t-2}^4C_{t-3}^4
+ 2C_{t-1}^2C_{t-2}^4C_{t-3}^2C_{t-4}^2
+ C_{t-1}^2C_{t-2}^2C_{t-3}^6 \cr
&\ \ \ \ \ \ \ \ \ \ \
+2C_{t-1}^2C_{t-2}^2C_{t-3}^4C_{t-4}^2 + C_{t-1}^2C_{t-2}^2
C_{t-3}^2C_{t-4}^4 \cr
&\ \ \ \ \ \ \ \ \ \ \
+ C_{t-1}^2C_{t-2}^2C_{t-3}^2
C_{t-4}^2C_{t-5}^2 \Bigr], &(AI.1f)\cr }$$
where $C_t=\cos \Bigl(k_{\perp} -t \phi \pi \Bigr)$.
\ These expressions can be seen to obey eq. (3.10a).

\vfill
\eject

\noindent {\bf REFERENCES}

\item{$^{\dagger}$} Present address: \ Dept. of Theoretical Physics,
University of Manchester, Manchester M13 9PL, UK.

\item{[1]} Hofstadter D H 1976 Phys. Rev. {\bf B 14} 2239

\item{[2]} Wannier G H 1978 Phys. Stat. Sol. (b) {\bf 88} 757

\item{[3]} Azbel M Ya 1964 Zh. Eksp. Theor. Fiz. {\bf 46} 929

\item{[4]} Freed D and Harvey J A 1990 Phys. Rev. {\bf B 41}
11328

\item{[5]} Wilkinson M 1984 Proc. Roy. Soc. London {\bf A 391}
305

\item{[6]} Rammal R and Bellissard J 1990 J. Phys. France {\bf 51}
1803

\item{[7]} Shapir Y and Wang X R 1990 Mod. Phys. Lett.
{\bf B 4} 1301

\item{[8]} Medina E, Kardar M, Shapir Y and Wang X R 1990
Phys. Rev. Lett. {\bf 84} 1916

\item{[9]} Shapir Y, Wang X R, Medina E and Kardar M 1990
in {\it Hopping and Related Phenomena}, ed H Fritzche
and M Pollak (World Scientific, Singapore, 1990)

\item{[10]} Fishman S, Shapir Y and Wang X R 1992 Phys. Rev.
{\bf B 46} 12154

\item{[11]} Affleck I and Marston J B 1988 Phys. Rev. {\bf B 37}
3774

\item{[12]} Nguyen V L, Spivak B Z and Shklovskii B I 1985
Pis'ma Zh. Eksp. Teor. Fiz. {\bf 41} 35 [JETP Lett. {\bf 41} 42]

\item{[13]} Zak J 1989 Phys. Lett. {\bf A 135} 385

\item{[14]} Entin-Wohlmam O, Haltzstein C and Imry Y 1986
Phys. Rev. {\bf B 34} 921

\item{[15]} Khveshchenko D V, Kogan Ya I, and Nechaev S K
1991 Int. J. Mod. Phys. {\bf B 5} 647

\vfill
\eject

\noindent{\bf Figure Captions}

\vskip 0.5truein

\noindent {\bf Figure 1.}  The diagonal lattice with $t$ and
$x$ axes labeled and an example of a directed path with backward
excursion.

\vskip 0.25truein

\noindent {\bf Figure 2.}  The Staggered Flux Configuration.

\vfill
\eject

\bye
\end